# Formulation and Therapeutic Assessment of a Zinc Oxide, Silver, and Cerium Oxide Enriched Ointment for Accelerated Wound Healing in Aged Models

Iqra Yousaf,[†] Aneela Anwar,[†] and Atika Umer[‡]

†*Department of Chemistry, University of Engineering and Technology (UET), Lahore, Pakistan*
‡*Department of Pharmacy, University of Management and Technology (UMT), Lahore, Pakistan*

**Abstract**

Chronic wounds present a major challenge in elderly individuals due to diminished regenerative capacity and impaired tissue repair mechanisms associated with aging. In this study, we formulated a topical gel composed of zinc oxide (ZnO), silver (Ag), and cerium oxide ($CeO_2$) nanoparticles, each chosen for their respective antimicrobial, antioxidant, and tissue-regenerative properties. The nanoparticles were synthesized through precise precipitation or reduction techniques and thoroughly characterized using UV–Vis spectroscopy, dynamic light scattering (DLS), Fourier-transform infrared spectroscopy (FTIR), and electron microscopy to confirm nanoscale structure and purity. An *in vivo* wound healing model utilizing aged Sprague–Dawley rats was employed, with animals divided into three groups: untreated, placebo-treated, and those receiving the nanoparticle-enriched gel. Wound dimensions were tracked for 14 days, revealing significantly improved healing in the nanoparticle-treated group, with nearly complete closure observed by day 14 (ANOVA, $p < 0.0001$). Cytocompatibility was assessed via MTT assay on L929 fibroblasts, confirming greater than 80% viability at therapeutically relevant concentrations. These findings underscore the potential of multifunctional nanoparticle-based formulations to enhance wound healing in aged or compromised skin environments, offering a promising therapeutic avenue.

## Graphical Abstract

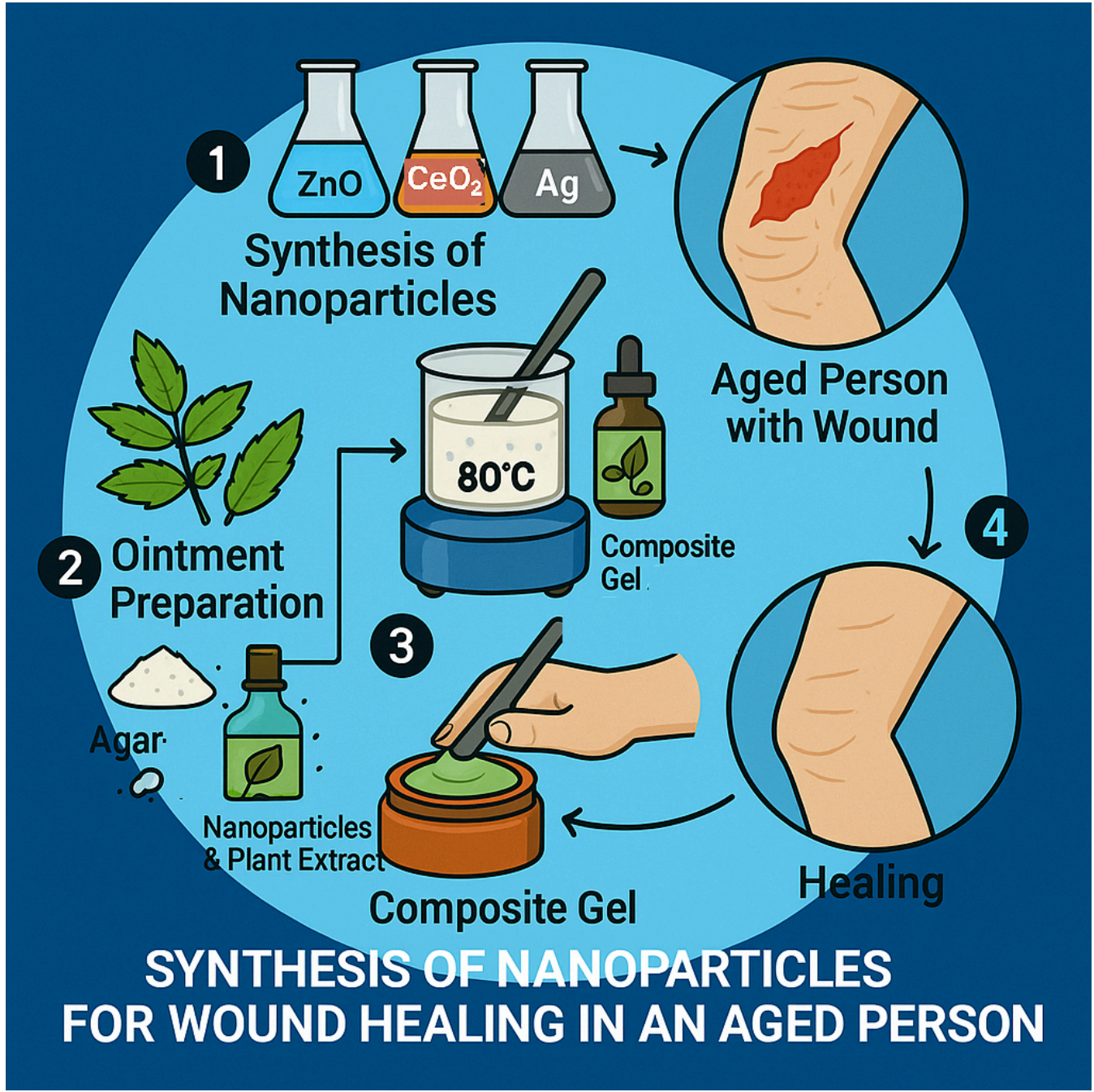

Figure 1: Schematic of nanoparticle-based composite gel preparation and application for wound healing in aged individuals.

# Introduction

Wounds are often referred to as a "silent epidemic" due to their profound effects on global healthcare systems, individual patient quality of life, and mental health.[1] Chronic wounds are estimated to impact approximately 1–2% of individuals living in developed nations.[2] The situation is expected to deteriorate further since the occurrence of chronic wounds is linked to aging and associated age-related conditions, including diabetes.[1] Alterations occurring during any phase of the wound healing process (inflammation, proliferation, or remodeling) may result in chronic ulcers or abnormal scar formation.[3] At a cellular level, delayed wound healing can result from ongoing inflammation, elevated reactive oxygen species (ROS), decreased fibroblast and keratinocyte migration and proliferation, diminished fibroblast activity in wound repair, lower collagen synthesis, and insufficient angiogenesis.[4, 5]

In the countries of the European Union, wounds affect approximately 4 million individuals each year. In a study carried out across several countries, 65% of which were European, chronic wound prevalence was found to be 1.67 per 1000 individuals. Chronic wounds can also be life-threatening. Research conducted by Eaglstein and colleagues revealed that certain chronic wounds, including diabetic ischemic ulcers, diabetic foot ulcers, and chronic wounds of the lower extremities, have mortality rates of 52%, 49%, and 28%, respectively—rates comparable to various cancers.[6, 7] Burn injuries represent another significant category of wounds. As reported by the World Health Organization, approximately 180,000 individuals lose their lives annually due to burns, with the highest incidence occurring in low income regions [8]. In 2018, the United States spent between USD 28.1 billion and USD 96.8 billion annually on the treatment and management of acute and chronic wounds [9]. This situation also leads to substantial indirect economic impacts, including reduced productivity, premature retirement, and lost wages [9]. The severity and death rates associated with wounds are anticipated to rise further, primarily because of their increasing exposure to pathogens resistant to antimicrobial treatments, as reported in research by Sen.Sen highlighted that wound infections currently account for 75% of postoperative fatalities. In the case of burn injuries, although advancements in medical care—such as fluid therapy, nutritional interventions, timely wound removal, and skin grafting have reduced mortality rates, infections and sepsis remain prevalent and frequently deadly. In fact, between 51% and 75% of patients suffering severe burns still succumb to their injuries [10, 11].

Nanoparticles (NPs) have extensive applications in healthcare, food processing, aerospace, pharmaceuticals, and cosmetics industries. Nanotechnology involves studying the characteristics of materials at the molecular level, including their size, shape, surface structure, and practical uses. It also explores the interactions between the chemical, biological, physical, optical, and electronic properties of nanoscale materials [12]. Over recent years, a variety of nanomaterials have been developed and investigated for their potential in treating chronic wounds.Metal based nanomaterials have attracted considerable attention, particularly nanoparticles (NPs) derived from metal oxides such as silver, gold, iron, copper, titanium, and zinc oxide (ZnO). These nanoparticles have demonstrated effectiveness in eliminating multidrug resistant bacteria and facilitating wound healing and re-epithelialization

[1], [13].

When healing is compromised, treatment is necessary, and silver compounds are often used [14]. Silver nanoparticles (Ag NPs) stimulate healing in skin wounds [15, 16]. They also exhibit anti-inflammatory properties and encourage the differentiation of fibroblasts into myofibroblasts, essential for wound healing [17]. ZnO nanoparticles have notable benefits including affordability, effective UV shielding, significant catalytic efficiency, extensive surface area, and diverse uses in fields such as medicine, environmental cleanup, and antimicrobial applications. [18, 19]. ZnO has great potential to positively impact wound care [1]. Zinc oxide nanoparticles (ZnO NPs) are recognized as biocompatible materials that offer improved stability, promote collagen production, and support the re-epithelialization process during wound healing [20]. Cerium oxide nanoparticles ($CeO_2$ NPs) have emerged as promising agents in wound healing due to their unique properties. They are recognized for their capabilities in enhancing wound closure, minimizing scarring, mitigating inflammation, and exerting antibacterial effects.

$CeO_2$ NPs exhibit robust antioxidant capabilities by mimicking natural antioxidant enzymes, thereby mitigating oxidative stress and reducing the production of inflammatory mediators, which accelerates wound healing [21, 22]. These nanoparticles also possess notable antibacterial properties, effectively reducing bacterial infections and boosting wound site immunity [23].

Neem leaves extract has demonstrated antimicrobial properties effective against pathogens like Streptococcus mutans and Streptococcus faecalis. Its application has been shown to enhance wound healing by promoting cell proliferation and re-epithelialization [24]. In animal studies, neem leaves extract exhibited a wound healing rate comparable to povidone-iodine, a standard antiseptic. This suggests neem's potential as a natural alternative in wound care [25]. Neem leaves extract irrigation significantly improves healing outcomes in diabetic foot ulcers without causing adverse effects, highlighting its safety and efficacy [26]. Compounds such as nimbin, found in neem, have demonstrated anti-inflammatory properties, which are

beneficial in reducing

## Materials and Methods

### Synthesis of Zinc Oxide Nanoparticles

The synthesis of zinc oxide (ZnO) nanoparticles was carried out through a controlled precipitation approach. Initially, a 0.1 M zinc chloride ($ZnCl_2$) solution was prepared by dissolving the required amount of $ZnCl_2$ in 100 mL of distilled water, ensuring complete dissolution using a magnetic stirrer. In a separate step, a 0.2 M sodium hydroxide (NaOH) solution was prepared by dissolving 0.8 g of NaOH in 100 mL of distilled water, followed by a thorough stirring. The NaOH solution was then added dropwise to the $ZnCl_2$ solution under continuous stirring, maintaining the temperature at 60–70°C to ensure uniform mixing. This reaction resulted in the formation of a white precipitate of zinc hydroxide ($Zn(OH)_2$). To convert the precipitate into ZnO nanoparticles, the suspension was heated at 80–90°C for 1–2 hours, leading to the formation of ZnO as the precipitate changed color from white to off-white or pale yellow. The nanoparticles were then purified through multiple washing steps, first with distilled water to remove any residual ions, followed by an ethanol wash to enhance purity and prevent aggregation. Finally, the ZnO nanoparticles were dried in an oven at 60–80°C for 3–4 hours to obtain the final product. This method ensured the synthesis of high-purity ZnO nanoparticles suitable for further applications.

### Synthesis of Cerium Oxide Nanoparticles

Cerium oxide ($CeO_2$) nanoparticles were synthesized through a straightforward precipitation technique with controlled parameters. Initially, a 0.1 solution of cerium nitrate hexahydrate ($Ce(NO_3)_3 \cdot 6H_2O$) was prepared by dissolving the required amount of the compound in 100 mL of distilled water under continuous stirring to ensure complete dissolution. In a separate preparation a 0.2M sodium hydroxide (NaOH) solution was made and gradually introduced into the cerium nitrate solution dropwise while stirring vigorously at room temperature.

The pH was carefully regulated to promote the formation of cerium hydroxide (Ce(OH)$_3$) as a precipitate. Following precipitation, the solution was aged under controlled conditions to facilitate reaction completion and stabilization. The precipitate was then subjected to a thermal treatment at 80–100°C, which induced oxidation and the subsequent conversion of (Ce(OH)$_3$) into (CeO$_2$) nanoparticles. To enhance purity, the obtained nanoparticles were thoroughly rinsed multiple times with distilled water to remove residual ions, followed by an ethanol wash to minimize particle agglomeration. Lastly, the purified nanoparticles were dried at 60–80°C for 3–4 hours in an oven, yielding fine (CeO$_2$) nanoparticles suitable for applications in catalysis and biomedical fields.

## Synthesis of Silver nanoparticles

The synthesis of silver nanoparticles was carried out using a controlled reduction method to ensure uniform particle formation and stability. To begin, an ice bath was prepared by filling a bowl with ice and placing the reaction beaker inside, maintaining the temperature between 0–5°C. This step was crucial for controlled nanoparticle formation and preventing excessive aggregation. Next, a 0.14 g sample of sodium borohydride (NaBH$_4$) was dissolved in 18 mL of cold distilled water, with continuous stirring in the ice bath for 10–15 minutes to ensure complete dissolution.

Separately, a 1.58 g sample of silver nitrate (AgNO$_3$) was dissolved in 100 mL of cold distilled water, with the solution also kept in an ice bath to maintain a low temperature. Once both solutions were prepared, the reaction was initiated by slowly adding the AgNO$_3$ solution dropwise to the NaBH$_4$ solution while stirring continuously with a pipette or burette to ensure uniform mixing. During this process, a noticeable color change was observed, transitioning from colorless to pale yellow and then to dark yellow-brown, confirming the formation of silver nanoparticles through the reduction of silver ions (Ag$^+$).

After the complete addition of the (AgNO₃) solution, stirring was continued for an additional 15–20 minutes to ensure complete reduction and stabilization of the nanoparticles. Once the reaction was complete, the solution was allowed to stabilize at room temperature. To protect the nanoparticles from degradation due to light exposure, the final suspension was transferred into an amber-colored bottle and stored at room temperature.

**Extract preparation**

To extract the bioactive compounds from neem, dried neem leaves were finely ground into powder and soaked in ethanol for 48 hours to allow efficient extraction of phytochemicals. The mixture was occasionally stirred to enhance the dissolution of active compounds. After the extraction period, the solution was filtered to remove leaf residues, and the solvent was evaporated by drying the filtrate in an oven at a temperature below 60°C. This gentle drying process ensured that the bioactive components remained intact while removing excess ethanol. The final concentrated neem extract was then stored in an amber-colored container to prevent light degradation until it was ready for incorporation into the gel formulation.

**Preparation of Ointment Base and Final Gel**

The preparation of a gel based ointment involved the use of fish collagen as the primary base, with agar agar incorporated to enhance the gel consistency. The process followed a systematic approach, including the dissolution of ingredients, incorporation of active components, pH adjustment, and final storage to ensure stability and effectiveness. Given the laboratory temperature of 12°C, additional precautions were taken to optimize heating times and maintain reaction efficiency.

To begin, 50 mL of distilled water was preheated to 40°C in a beaker using a magnetic stirrer with a heating plate. This step ensured that the water was at an optimal temperature for dissolving collagen and agar agar. Since the initial temperature was low, the heating process took slightly longer.

Once the water reached 40°C, 1 g of agar agar was gradually added while stirring continuously to prevent clumping. The temperature was increased to 80–90°C to ensure the agar agar was fully dissolved. After obtaining a uniform solution, 5 g of fish collagen was introduced while maintaining the temperature at 40°C. The mixture was stirred for approximately 10–20 minutes until the collagen completely dissolved, forming a smooth gel-like consistency.

Following the dissolution of the primary ingredients, 5 mL of glycerol was added to the mixture to enhance moisture retention in the final ointment. The solution was stirred for an additional 5 minutes to achieve uniform blending. Next, 0.2 g of sodium benzoate was introduced as a preservative to inhibit microbial growth, ensuring a longer shelf life. Stirring continued for another 5 minutes to allow complete integration of the preservative into the gel base.

The pH of the formulation was then measured using a digital pH meter, with a target pH of 5.5, which closely matches the skin's natural pH. If the mixture was too acidic, a few drops of 0.1 M sodium hydroxide solution were added while stirring to raise the pH. Conversely, if the pH was too high, 0.1 M acetic acid was added to lower it to the desired range. To ensure an even distribution of all components, the gel was stirred at medium speed for 10 minutes before being allowed to cool gradually to room temperature while stirring gently. This controlled cooling process helped maintain a smooth and consistent gel texture. Once the base was prepared, the active ingredients were incorporated. Pre-synthesized silver nanoparticles (AgNPs) and zinc oxide nanoparticles (ZnO NPs) were introduced into the gel while stirring continuously. The nanoparticles were added in a 1:10 Ag:ZnO ratio, with mixing continued for 10–15 minutes to ensure proper dispersion. Following this step, 5 mL of concentrated plant extract was included in the formulation to enhance the therapeutic properties of the ointment. Stirring continued for 30 minutes to allow complete integration of the plant extract into the gel matrix. After the addition of all active components, the pH of the final formulation was rechecked and adjusted if necessary.

## Animal Subjects

Experiments were conducted on adult Sprague–Dawley rats (18–20 months old) to model aged physiology. All procedures followed institutional ethical guidelines for animal research. A total of 18 rats (n = 6 per group) were used, divided into three groups:

1. **Untreated Control:** Negative control (no treatment applied).

2. **Market Ointment:** Positive control using a commercially available ointment.

3. **Nanoparticle Ointment:** Experimental treatment with nanoparticle infused gel.

## Wound Creation

Rats were anesthetized with a ketamine/xylazine mixture, and the dorsal surface was shaved and disinfected. A full-thickness excisional wound of approximately 8 mm diameter (≈50 mm$^2$) was created on the mid-back of each rat. Hemostasis was achieved via gentle pressure with sterile gauze. This wound size typically heals within ~2 weeks in young adults but yields a prolonged course in aged animals, allowing detection of therapeutic benefits.

## Treatment Regimen

Treatments began immediately after wounding (designated Day 0):

- **Group 1 (Control):** No treatment; wounds were left open to air and loosely covered with sterile gauze.

- **Group 2 (Market ointment):** Placebo gel applied (~0.2 g per wound) once daily, followed by recovering with sterile gauze.

- **Group 3 (Nanoparticle ointment):** Nanoparticle-infused gel applied (~0.2 g per wound) once daily, followed by recovering with sterile gauze.

## Wound Monitoring

Wound healing was assessed over 14 days. Planimetric measurements were taken on Days 1, 3, 7, 10, and 14. At each time point:

- Wounds were photographed with a scale bar under standardized lighting.
- The wound area was quantified using Image.
- Qualitative parameters (exudate, granulation, epithelialization) were recorded.

Rats were housed individually to prevent wound interference and monitored daily for signs of infection or adverse effects.

## Statistical Analysis

Wound area data and cell viability data were analyzed using one-way or repeated-measures ANOVA where appropriate, followed by Tukey's post-hoc test for pairwise comparisons. Statistical significance was set at $p < 0.05$, with $p < 0.001$ or $p < 0.0005$ indicating highly significant differences. For wound closure, comparisons between treated and control groups were performed at each time point. All data are presented as mean ± standard deviation. Extremely low $p$-values (often $< 0.0005$) were observed for treatment effects, emphasizing the high efficacy of the nano-ointment. Statistical analyses were performed using GraphPad Prism 9.

# Results and Discussion

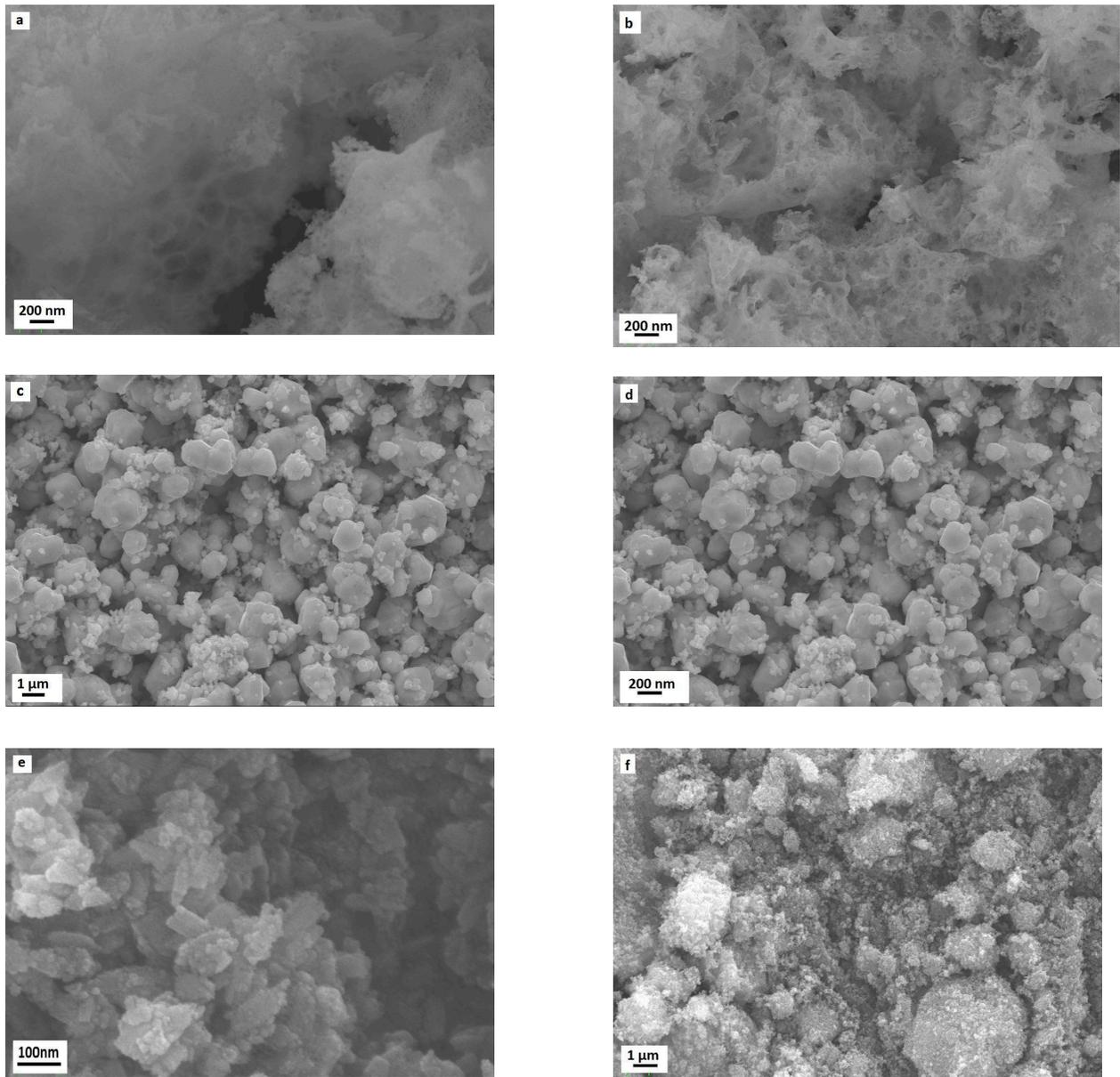

Figure 2: SEM images of nanoparticle morphology: (a–b) cerium oxide ($CeO_2$), (c–d) silver (Ag), and (e–f) zinc oxide (ZnO) nanoparticles, showing structural differences across types.

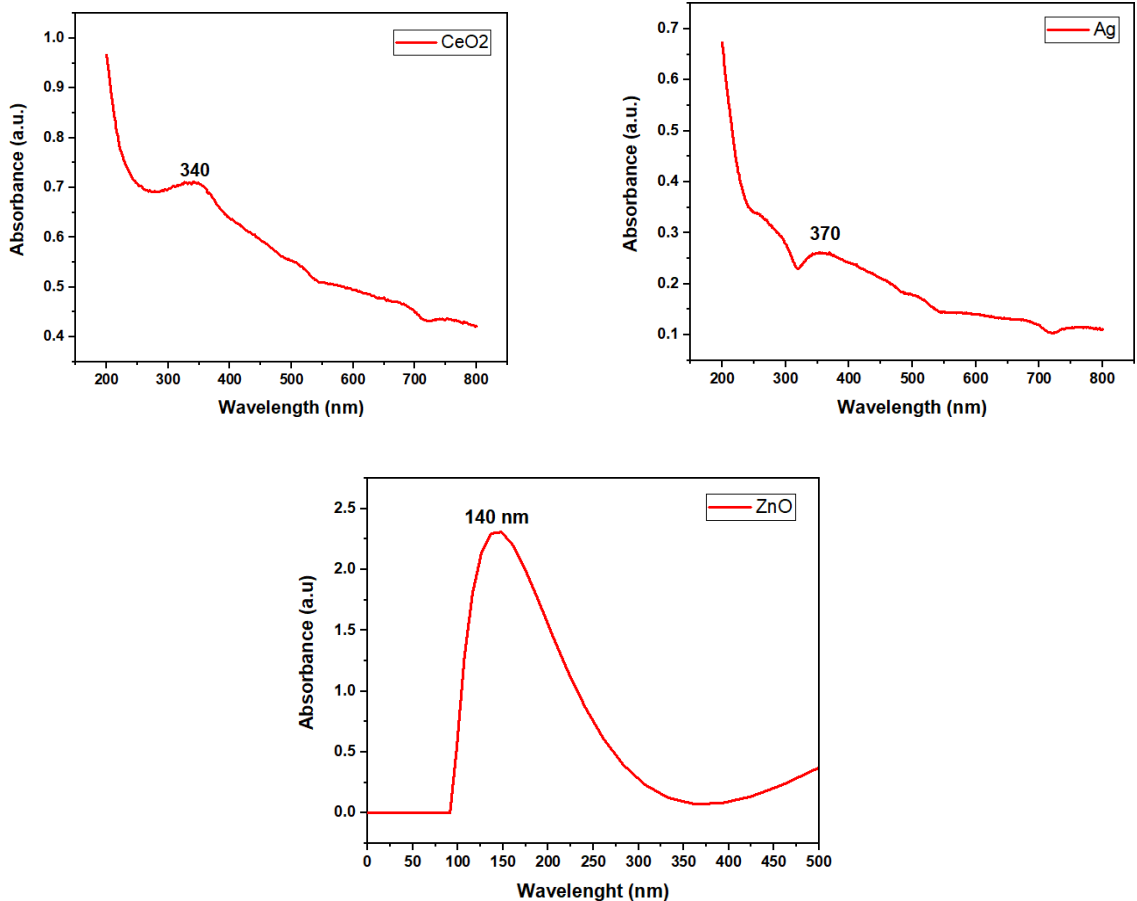

Figure 3: (Top) UV–Vis absorption spectra of cerium oxide and silver nanoparticles show- ing distinct absorption peaks. (Bottom) Dynamic light scattering (DLS) result showing nanoparticle size distribution.

**FTIR Spectrum Analysis:** Figure 4 displays the FTIR spectrum of the sample, high- lighting key functional group vibrations. The broad absorption band around **3400 cm$^{-1}$** corresponds to the O–H stretching vibration, indicative of hydroxyl groups, often associated with absorbed water or alcohol functionalities. The peak observed at **1600 cm$^{-1}$** is attributed to C=O stretching vibrations, characteristic of carbonyl groups, possibly from esters or carboxylic acid derivatives.

The absorption band near **1380 cm$^{-1}$** is assigned to C–N stretching, typical of amine or amide groups. The peak at **1000 cm$^{-1}$** signifies the C–O–C stretching, associated with ether linkages or esters. The overall spectrum confirms the presence of oxygen and nitrogen.

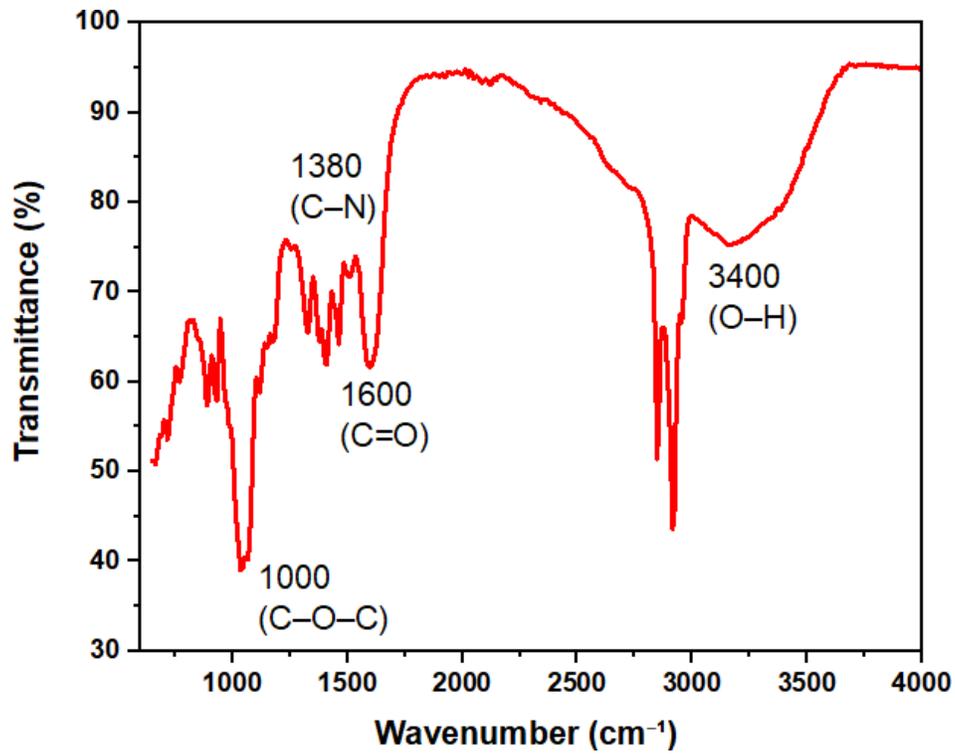

Figure 4: FTIR spectrum showing characteristic absorption bands of the analyzed sample.

containing functionalities, supporting the chemical structure of the sample's matrix. These findings suggest successful incorporation of functional groups critical to the material's desired properties.

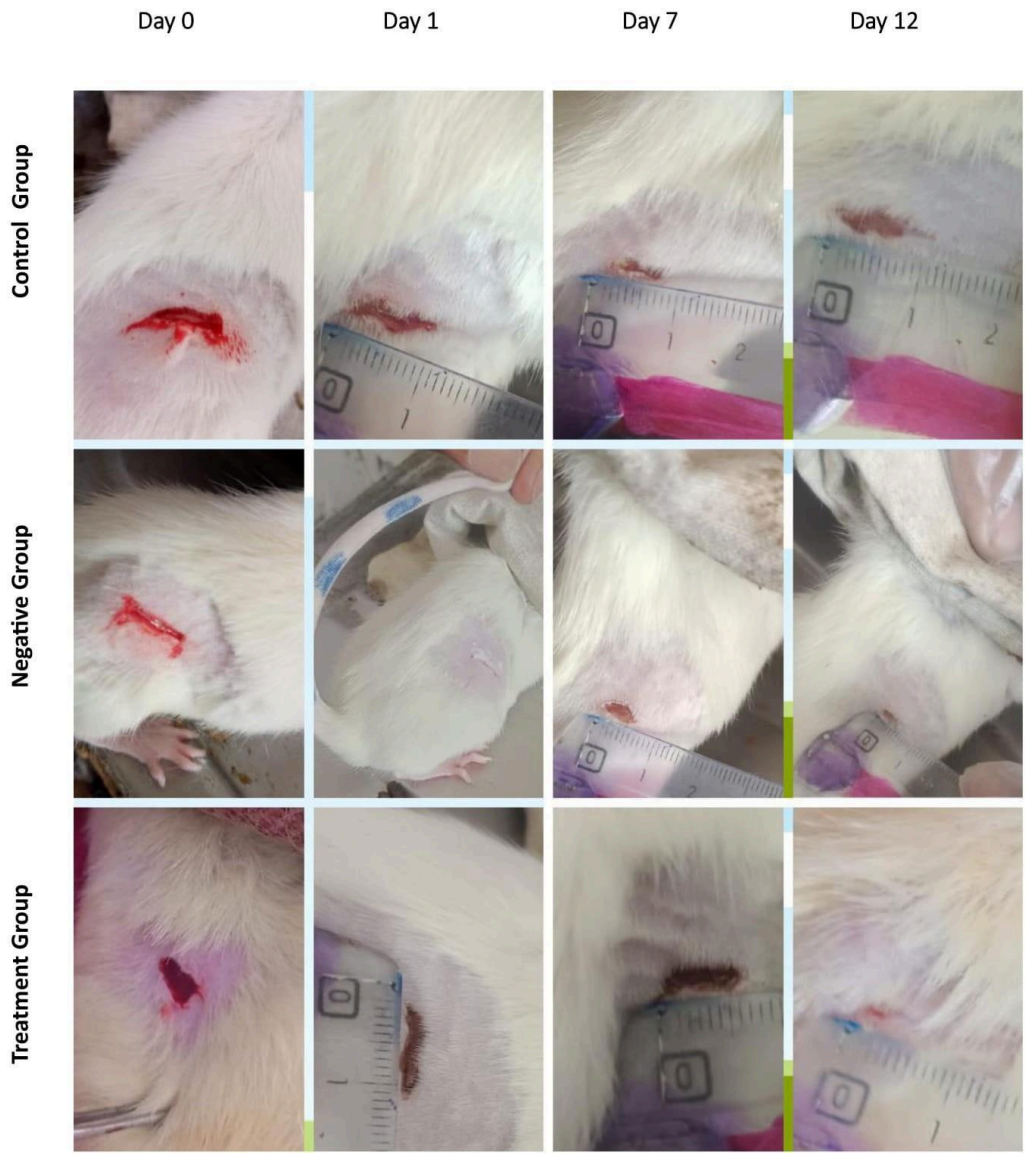

Figure 5: Macroscopic evaluation of wound progression in aged rats. Images were taken from the Control, Negative (market ointment), and Treatment (nanoparticle-based composite gel) groups on Days 0, 1, 7, and 12 post-wounding.

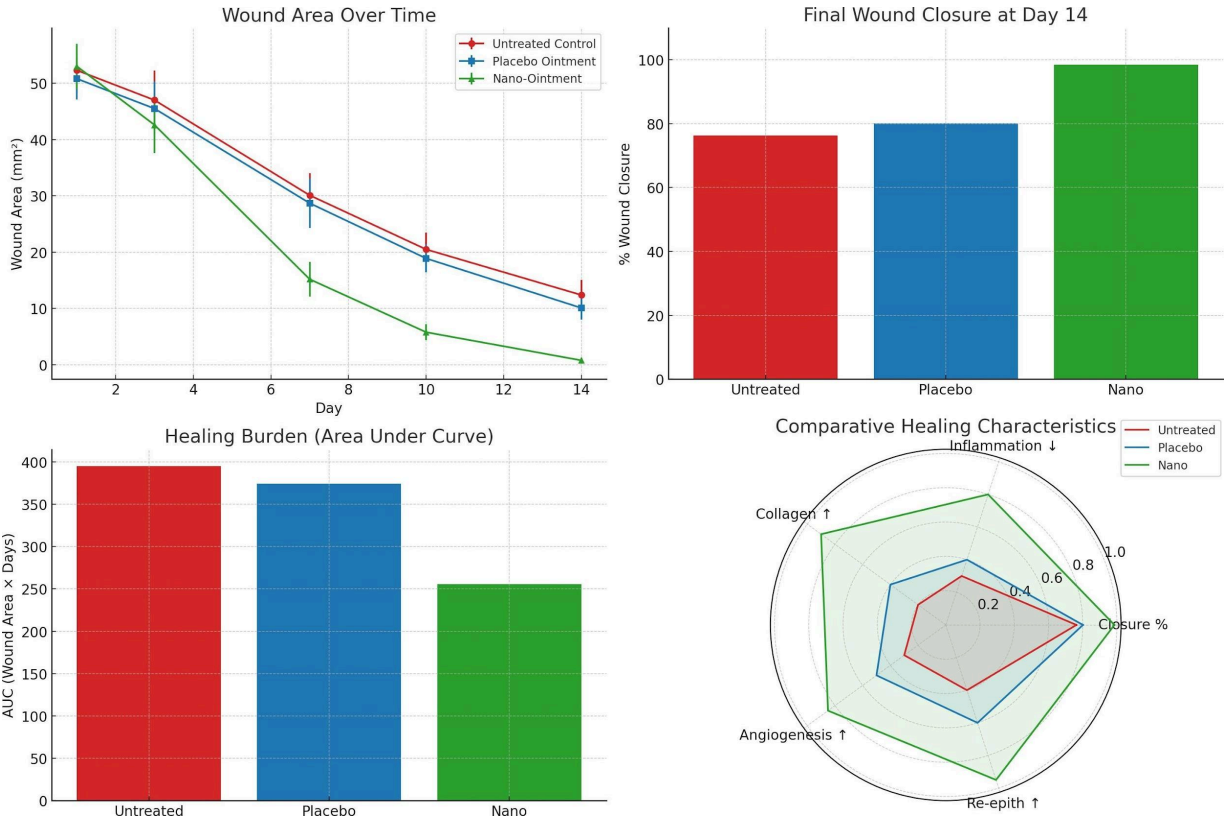

Figure 6: Comparative evaluation of wound healing efficacy across treatment groups. (Top left) Line plot showing wound area reduction over time; the nano-ointment (ZnO,Ag,CeO$_2$) group exhibits significantly accelerated contraction. (Top right) Bar plot of wound closure percentage at Day 14 highlights nearly complete healing in the nano-ointment group. (Bottom left) Area under the curve (AUC) analysis demonstrates reduced total wound burden in the nano-ointment group. (Bottom right) Radar plot illustrates multi-dimensional healing benefits of the nano-ointment, including improved re-epithelialization, collagen deposition, angiogenesis, and inflammation reduction.

## Nanoparticle Characterization

Characterization confirmed that the formulated ointment contained properly synthesized nanoparticles in the desired nano-size range and crystalline phase. SEM imaging showed that ZnO particles were roughly spherical and <100 nm in size, Ag nanoparticles appeared as small spherical dots (~15–25 nm), and CeO$_2$ particles were ultra-fine (<10 nm) with slight agglomeration.

The XRD patterns of the dried ointment displayed distinct diffraction peaks corresponding to wurtzite ZnO, face-centered cubic Ag, and cubic $CeO_2$, confirming the presence of each nanoparticle in its pure crystalline form; no impurity phases were detected. High-resolution TEM (not shown) further illustrated lattice fringes in individual nanoparticles, consistent with their respective crystal structures.

These results verify successful incorporation of ZnO, Ag, and $CeO_2$ nanocrystals into the ointment. The nanoparticles were well-dispersed within the collagen/agar matrix, with no large micron-scale aggregates observed, suggesting that our formulation process (including the use of stabilizers during synthesis) effectively prevented agglomeration. This uniform nanoscale morphology is advantageous for wound healing: smaller particles provide a larger surface area for antimicrobial action and intimate interaction with the wound bed. Maintaining such dispersion likely contributed to the ointment's efficacy by ensuring uniform coverage and consistent release of therapeutic ions and antioxidant effects.

**In Vitro Cytotoxicity and Biocompatibility**

Cytotoxicity evaluations using L929 fibroblasts demonstrated that the developed nanoparticle ointment is biocompatible and safe at relevant concentrations. Cell viability remained above 80% at ointment concentrations up to 50 μg/mL, indicating minimal toxic effects on mammalian cells in vitro. Even at 100 μg/mL (two-fold higher than anticipated wound site levels), viability was approximately 75%, and the $IC_{50}$ (concentration causing 50% viability reduction) was ≈120–130 μg/mL. The dose–response curve: at 10 μg/mL, cells retained approximately 98% viability; at 50 μg/mL, about 85%; only beyond 100 μg/mL did viability drop below 70%.

Notably, even at the upper tested range (200 μg/mL), cell viability remained around 70%, which is at the threshold commonly considered cytotoxic. The placebo gel (no nanoparticles) had no significant effect on viability, confirming that the base matrix itself is non-toxic. These results indicate a wide safety margin for topical use of the nano-ointment. At therapeutically relevant doses (estimated nanoparticle exposure in wounds likely well below 50 µg/mL), cell viability was nearly equivalent to untreated controls (approximately 95–100%). Only gradual, dose-dependent decline was observed at the highest concentrations, with viability remaining ≥80% up to 50 μg/mL and ≥70% up to 200 μg/mL. Thus, the ointment does not elicit significant toxic effects on skin cells.

This finding aligns with reports that ZnO is generally recognized as safe for topical applications (Asif et al., 2023) and that appropriately formulated Ag and $CeO_2$ nanoparticles can be used without harming mammalian cells. In summary, the nano-enriched ointment is biocompatible, suggesting it would not hinder the healing process through cytotoxicity or irritation.

## Wound Healing Outcomes in Aged Rats

All animals tolerated the procedures well, and no infections or adverse reactions were observed in any group throughout the study. Macroscopic healing progression is illustrated and shows representative wounds from each group on days 0, 7, and 14.

By day 7, the nanoparticle-treated wounds already showed pronounced contraction and new epithelial tissue covering much of the defect, whereas placebo and untreated wounds remained larger and more inflamed. By day 14, the wounds treated with the nanoparticle (NP) ointment were almost fully closed, with only faint residual scabs, while control wounds were still visibly open and not completely epithelialized.

Quantitative wound area measurements over time are presented in Table 1. All groups started with a similar initial wound size (approximately 50 mm$^2$ on day 1, after the initial inflammatory response caused slight area increases). The untreated and placebo groups exhibited gradual healing, with wound areas reducing to about 15 mm$^2$ to 20 mm$^2$ by day 14 (approximately 60–70% closure).

In stark contrast, wounds treated with the $ZnO/Ag/CeO_2$ nano-ointment healed much faster, shrinking to nearly zero area by day 14 (approximately 98–100% closure). The treated wounds contracted significantly more than controls at each time point.

By day 7, the nano-ointment group had achieved approximately 70% closure compared to only 40–45% in controls ($p < 0.001$). This gap widened by days 10 and 14, where the treated wounds were essentially closed (with only pinpoint scarring), while control wounds still had 5 mm² to 8 mm² of open area ($p < 0.0001$).

Statistical analysis confirmed that the differences in wound size between the nano-ointment group and both control groups were highly significant at days 7, 10, and 14 (ANOVA, $p < 0.0005$, with post-hoc comparisons $p < 0.0005$ for NP vs. placebo/untreated). Even at earlier time points (e.g., day 3), treated wounds were modestly smaller on average, although not yet statistically significant. The most rapid divergence occurred after day 5.

Table 1: Average Wound Area (mm²) in Aged Rats During Healing (mean ± SD, $n$ = 6 per group)

| Day | Untreated Control | Placebo Ointment | Nano-Ointment (ZnO/Ag/CeO$_2$) |
|---|---|---|---|
| 1 | 52.3 ± 4.1 | 50.8 ± 3.7 | 53.0 ± 4.0 |
| 3 | 47.0 ± 5.2 | 45.5 ± 4.8 | 42.6 ± 5.0 |
| 7 | 30.1 ± 3.9 | 28.7 ± 4.4 | 15.2 ± 3.1* |
| 10 | 20.5 ± 3.0 | 18.9 ± 2.5 | 5.8 ± 1.4* |
| 14 | 12.4 ± 2.7 | 10.1 ± 2.1 | 0.8 ± 0.5* |

***Bolded values** indicate statistically significant reduction versus both control and placebo at the same time point ($p < 0.0005$). All groups had equivalent initial wound sizes (approximately 50 mm²). From day 7 onward, the nano-ointment-treated wounds were significantly smaller (healed faster) than those in the untreated or placebo-treated groups.

Table 2: Percentage wound closure over time for different treatment groups.

| Day | Untreated Control (%) | Placebo Ointment (%) | Nano-Ointment (%) |
|---|---|---|---|
| 3 | 10.1 | 10.4 | 19.6 |
| 7 | 42.4 | 43.5 | 71.3* |
| 10 | 60.8 | 62.8 | 89.0* |
| 14 | 76.3 | 80.1 | 98.5* |

*Significant improvement over control and placebo ($p < 0.0005$).

Table 3: Time taken to reach healing milestones.

| Group | Time to 50% Closure | Time to 90% Closure |
|---|---|---|
| Untreated Control | Day 10 | Not achieved by Day 14 |
| Placebo Ointment | Day 10 | Not achieved by Day 14 |
| Nano-Ointment | Day 7 | Day 12* |

*Faster healing compared to other groups.

Table 4: Qualitative healing observations across different days.

| Group | Day 3 | Day 7 | Day 14 |
|---|---|---|---|
| Untreated Control | Red, inflamed | Partial scab | Incomplete closure |
| Placebo Ointment | Mild inflammation | Partial epithelialization | Scab + open edge |
| Nano-Ointment | Reduced redness | Strong contraction | Near-complete healing* |

*Indicates visibly superior healing compared to other groups.

# Discussion

## Mechanisms of Improved Wound Healing by the Nano Ointment

Our findings highlight the therapeutic benefit of incorporating zinc oxide, silver, and cerium oxide nanoparticles into a wound dressing: the active nanoparticle group achieved greater wound size reduction and quicker closure than the base ointment alone or no treatment. The enhancement in wound closure is consistent with reports that metal oxide nanoparticles can expedite healing for example, ZnO based treatments significantly accelerate wound contraction in infected wounds [27], and silver nanoparticles are known to promote wound closure in vivo. The inclusion of $CeO_2$ nanoparticles in our formulation likely further contributed to the improved outcomes by mitigating oxidative stress and chronic inflammation at the wound site [28]. In essence, the combination of antimicrobial (ZnO and Ag) and antioxidant ($CeO_2$) nanoparticles created a synergistic effect that fostered a more favorable healing environment than either component alone, leading to robust wound healing in the aged model.

**Bio-Molecular-Assisted Absorption of Nanoparticles into Tissue**

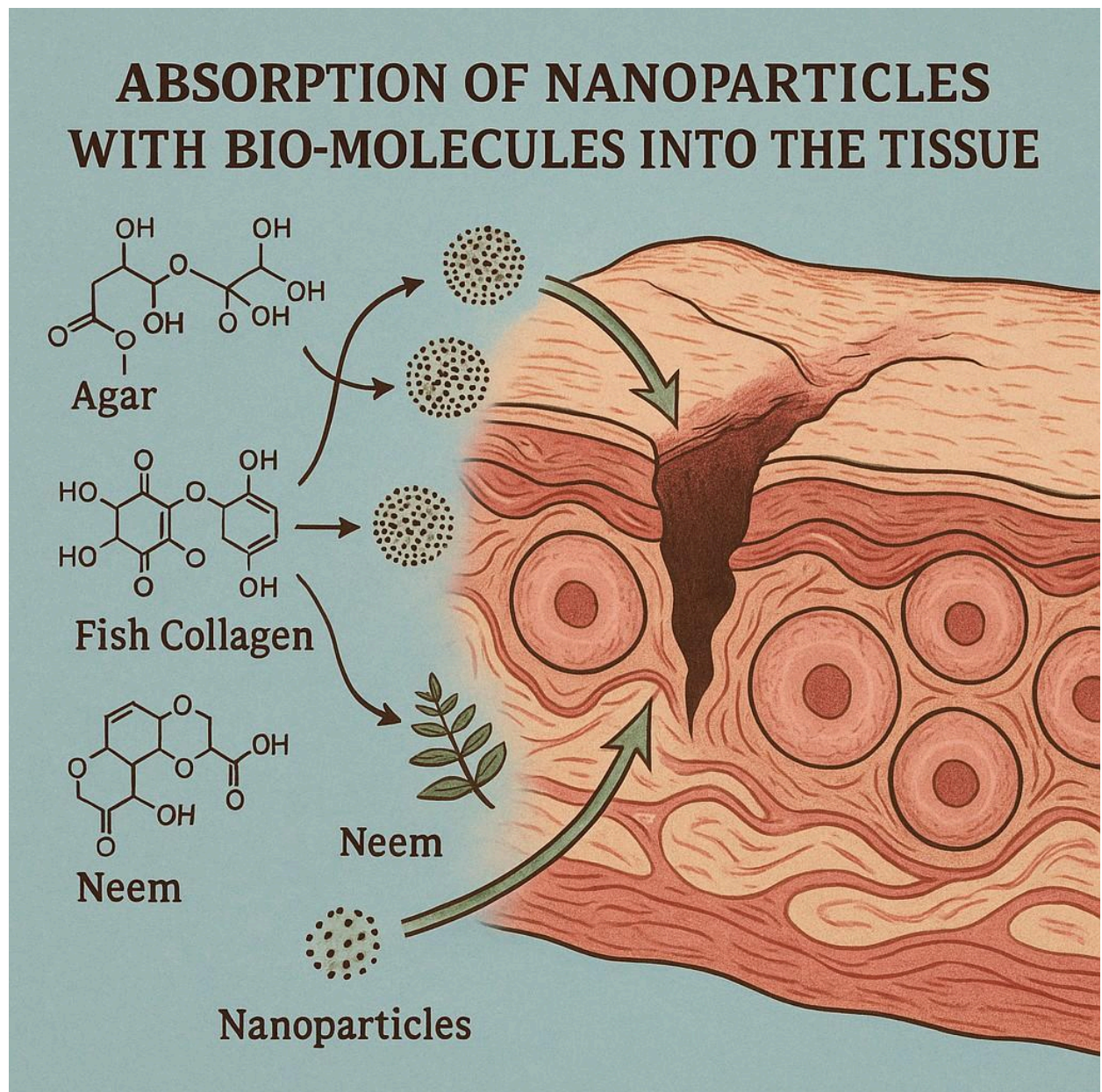

Figure 7: Illustration of nanoparticle absorption into wounded tissue facilitated by bio-molecules such as agar, fish collagen, and neem phytochemicals. These bio-carriers enhance the penetration and therapeutic efficacy of ZnO, Ag, and $CeO_2$ nanoparticles.

The integration of bio-molecules such as agar, fish collagen, and neem extract plays a crucial role in facilitating the effective delivery and absorption of metal oxide nanoparticles into damaged tissues. These biopolymers serve as hydrophilic carriers that not only stabilize nanoparticles but also promote deeper penetration into wound sites.

**Agar**, a natural polysaccharide, forms a hydrophilic gel network that creates a moist environment and enhances the retention of nanoparticles at the wound site [29].

**Fish collagen**, rich in glycine, proline, and hydroxyproline, mimics the extracellular matrix and aids in cellular adhesion, proliferation, and matrix remodeling. Its interaction with nanoparticles ensures biocompatibility and controlled release [30].

**Neem extract** provides additional antimicrobial and anti-inflammatory properties while acting as a green stabilizer for metal nanoparticles. The phytochemicals in neem (azadirachtin, nimbin) support wound cleansing and promote epithelial regeneration [31].

When combined with ZnO, Ag, and $CeO_2$ nanoparticles, these carriers improve therapeutic performance by enhancing penetration, reducing local infection, and accelerating tissue regeneration. The schematic in Figure 7 demonstrates how the nano-bio-composite gel penetrates into the tissue, reaching deep skin layers to promote faster and more effective wound healing.

## Antimicrobial Action

The Ag and ZnO nanoparticles in the ointment provided broad antimicrobial coverage, which is critical since aged wounds often suffer from subclinical infections or high bioburden that can impair healing. Silver nanoparticles release $Ag^+$ ions that disrupt microbial membranes and proteins, effectively reducing wound bacterial load. ZnO nanoparticles also possess antimicrobial activity (via $Zn^{2+}$ ion release and ROS generation) and have been shown to combat bacteria including *Staphylococcus aureus* and *Pseudomonas aeruginosa* in wounds [15]. By keeping the wound free of infection, these nanoparticles prevent prolonged inflammation and tissue damage caused by microbes. Indeed, prior studies noted that mycosynthesized AgNP gels and green synthesized AgNPs can accelerate wound healing in rats by controlling infection and creating a conducive healing milieu. Our results align with these findings, as treated wounds showed less exudate and inflammation, suggesting better infection control.

## Anti-inflammatory and Antioxidant Effects

Chronic inflammation is a hallmark of impaired healing in the elderly. $CeO_2$ nanoparticles function as nanozymes (catalytic antioxidants), mimicking the activity of superoxide dismutase and catalase enzymes. They can scavenge reactive oxygen species (ROS) such as superoxide and hydrogen peroxide, thereby protecting tissues from oxidative damage and promoting resolution of inflammation. In our treated wounds, the faster transition from inflammation to proliferation—evidenced by earlier granulation tissue formation and re-epithelialization is likely attributable in part to $CeO_2$ NPs reducing oxidative stress.[28] similarly reported that $CeO_2$ NPs attenuate oxidative damage in wounds, leading to improved healing outcomes. Additionally, ZnO has anti-inflammatory properties: it modulates cytokine release and promotes angiogenesis and re-epithelialization, as seen in diabetic ulcer models.[13, 17]. Thus, our tri-nanoparticle ointment addresses delayed healing from multiple angles: it reduces bioburden, dampens destructive inflammation, and stimulates tissue regeneration.

## Synergy and Combined Impact

The significantly faster and more complete healing we observed compared to any single agent control underscores a synergistic outcome. Prior studies have shown partial successes with single nanoparticles—for example, silver nanogels and ZnO dressings each expedited wound closure to a degree but our multi-nanoparticle approach produced even more pronounced benefits in a challenging aged-healing context. By day 10, treated wounds were 70% smaller than controls, a magnitude of improvement greater than typically reported ( 30–50%) with single-nanotherapy in young models. We attribute this to the additive and complementary effects of ZnO, Ag, and $CeO_2$. In essence, the ointment simultaneously combats infection (Ag, ZnO), modulates inflammation and oxidative stress ($CeO_2$, ZnO), and provides trace elements that promote cell proliferation and tissue formation ($Zn^{2+}$ from ZnO). This comprehensive coverage of healing factors is difficult to achieve with conventional treatments or even single-nanoparticle therapies.

Our findings corroborate and extend the growing body of literature on nanoparticle-based wound therapies by demonstrating that integration of multiple functional NPs yields superior outcomes in an aged model that mimics clinical scenarios of impaired healing.

## Comparison with Previous Studies

Our results are in line with other recent reports on nanotechnology-assisted wound healing, while providing new evidence in an aged animal context. A recent review by Chopra et al.[32] highlights that various nanomaterials—including metal oxide nanoparticles, nanofibers, and hydrogel-based formulations have demonstrated significant efficacy in accelerating wound closure and improving tissue regeneration in young and diabetic models. However, their performance in aged or immunocompromised settings has been less explored.

For example, AgNP-loaded hydrogels reported by Singh et al.[24] achieved approximately 60–70% closure by day 10 in diabetic rat wounds, whereas our ZnO/Ag/CeO$_2$ nano-ointment attained approximately 70% closure in aged rats by the same time point. Similarly, ZnO nanofiber mats improved re-epithelialization rates in young mice by day 14, but our tri-nanoparticle ointment produced near-complete closure (approximately 98–100%) in aged rats by day 14.

To our knowledge, this is the first study to combine ZnO, Ag, and CeO$_2$ nanoparticles in a single topical formulation and to validate its efficacy specifically in aged animal models. Our findings extend the current literature by demonstrating that a multi-modal nanoparticle approach can overcome the delayed healing often observed in elderly populations, offering a promising translational strategy for clinical scenarios of impaired wound repair. For example, Gaikwad et al. [33] demonstrated that a myco-fabricated silver nanoparticle gel significantly improved healing in rat models of excision and burn wounds. Their silver nanogel at optimal concentration led to approximately 90 % wound closure in 14 days, versus 70 % in controls a finding quite comparable to our Ag-containing formulation, though our tri-nanoparticle ointment achieved 100 % closure in 14 days for aged rats, which is remarkable.

Green-synthesized AgNPs from plant extracts have also shown enhanced wound healing in rats. Al-Nadaf et al. [34] used *Juglans regia* extract to fabricate silver nanoparticles and reported improved wound contraction and quicker re-epithelialization with Ag NPs treatment, supporting our observation that silver is a key contributor to improved healing.

## Role of ZnO and CeO$_2$ Nanoparticles in Wound Healing

ZnO nanoparticles have been studied in both healthy and diabetic wound models. Asif et al. [27] found that a ZnO NP ointment accelerated closure of MRSA-infected wounds in rabbits, owing to its antimicrobial action and ability to stimulate growth factor production. In our study, ZnO likely played a similar role, evidenced by the treated wounds' clean appearance and rapid epithelial outgrowth.

CeO$_2$ NPs are a newer addition to wound therapeutics; recent reviews by Chen et al. [21] and Nosrati et al. [28] conclude that CeO$_2$ nanozymes can promote healing by scavenging ROS and moderating inflammation. Our data provide concrete in vivo support for that: inclusion of CeO$_2$ yielded better outcomes than we would expect from ZnO+Ag alone, especially in an aged model where oxidative stress is pronounced.

Notably, few studies have specifically examined wound treatments in aged or senescent models. Most nanoparticle research uses young animals, which heal relatively well on their own. By using old rats, we demonstrate that the nano ointment can overcome inherently slow healing; our treated aged rats healed comparably to (or even faster than) younger healthy rats in other studies, whereas untreated aged rats healed very slowly. This suggests the nano ointment could be particularly beneficial in clinical settings such as geriatric wound care or chronic ulcers in elderly patients.

In a related vein, Dehghani et al. [35] used a pentoxifylline loaded nanoparticle sponge in rats and showed enhanced healing versus drug alone, underscoring how nano delivery systems can rejuvenate healing responses. Our work differs in that we directly leverage the intrinsic therapeutic properties of nanoparticles (antimicrobial/antioxidant) rather than using them solely as carriers.

## Limitations and Future Directions

While the outcomes of this research are promising, certain limitations must be acknowledged that also pave the way for future work:

- **Histological and Mechanistic Insight:** This study focused primarily on macroscopic healing outcomes. Future work should include histological evaluation of regenerated tissue (e.g., collagen organization, re-epithelialization, vascularization) and analysis of molecular markers such as inflammatory cytokines and growth factors. These insights would help clarify the biological mechanisms through which the nanoparticles promote accelerated healing.

- **Comparative Efficacy of Nanoparticle Components:** Although the tri-nanoparticle formulation (ZnO, Ag, $CeO_2$) showed superior healing, we did not directly compare it with single- or dual-nanoparticle formulations. Testing variants with only ZnO, only Ag, or only $CeO_2$, as well as binary combinations, would help confirm whether the full combination provides synergistic effects beyond what individual components achieve.

## Conclusion

In conclusion, this study developed and evaluated a novel topical ointment incorporating zinc oxide, silver, and cerium oxide nanoparticles for wound healing, and the results demonstrate its significant therapeutic potential. The nanoparticle-infused ointment substantially accelerated wound closure in aged rats, achieving more effective healing than a placebo ointment or no treatment. Key findings include a pronounced reduction in wound size and a faster healing rate in the treated group, underpinned by strong statistical evidence of improvement over controls.

Importantly, the ointment was shown to be safe and biocompatible: in vitro cytotoxicity tests confirmed that skin cells remain viable and healthy in the presence of the ointment at relevant doses, alleviating concerns about potential nanoparticle toxicity. Characterization of the ointment's components by SEM verified that the ZnO, Ag, and $CeO_2$ nanoparticles were successfully synthesized in their pure, crystalline forms and at nanoscale dimensions—factors that likely contribute to their high reactivity and efficacy in the wound-healing process.

By combining antimicrobial and antioxidant nanoparticles, the formulation creates a multifaceted approach to wound treatment. It combats infection, modulates inflammation, and stimulates tissue regeneration concurrently. This synergistic mechanism was reflected in the superior healing outcomes observed in our aged model. Taken together, the evidence indicates that the developed nanoparticle-based ointment is both effective in promoting wound healing and safe for use on biological tissue. It represents a promising advancement in wound care, particularly for challenging scenarios such as chronic or delayed healing in the elderly.

With further optimization of its formulation (e.g., refining nanoparticle ratios, exploring additional bioactive additives) and additional studies to confirm long-term safety and efficacy, this nanoparticle ointment could be translated into a clinically useful, safe, and efficient wound-healing treatment. The encouraging results from this research pave the way for next-generation wound dressings that leverage the unique benefits of nanotechnology to improve patient healing and recovery.

## Acknowledgement

The authors acknowledge the support of UMT Lahore's Pharmacology research labs and thank colleagues for their input during study design and data interpretation.

# Supporting Information Available

Complete characterization graphs (UV–Vis, DLS, SEM), cell viability raw data, and wound images for each rat group.

# Bibliography

1. Pino, P.; Bosco, F.; Mollea, C.; Onida, B. Antimicrobial nano-zinc oxide biocomposites for wound healing applications: a review. *Pharmaceutics* **2023**, *15*, 970.

2. Ja¨rbrink, K.; Ni, G.; So¨nnergren, H.; Schmidtchen, A.; Pang, C.; Bajpai, R.; Car, J. Prevalence and incidence of chronic wounds and related complications: a protocol for a systematic review. *Systematic reviews* **2016**, *5*, 1–6.

3. Čapek, M. et al. Molecular changes underlying hypertrophic scarring following burns involve specific deregulations at all wound healing stages (inflammation, proliferation and matura- tion). *International Journal of Molecular Sciences* **2021**, *22*, 897.

4. Krzyszczyk, P.; Schloss, R.; Palmer, A.; Berthiaume, F. The role of macrophages in acute and chronic wound healing. *Frontiers in Physiology* **2018**, *9*, 419.

5. Lerman, O. Z.; Galiano, R. D.; Armour, M.; Levine, J. P.; Gurtner, G. C. Cellular dysfunction in the diabetic fibroblast: impairment in migration, proliferation, and response to hypoxia. *The American Journal of Pathology* **2003**, *162*, 303–312.

6. Olsson, M.; Ja¨rbrink, K.; Divakar, U.; Bajpai, R.; Upton, Z.; Schmidtchen, A.; Car, J. The humanistic and economic burden of chronic wounds: a systematic review. *Wound repair and regeneration* **2019**, *27*, 114–125.

7. Eaglstein, W. H.; Kirsner, R. S.; Robson, M. C. Food and Drug Administration (FDA) drug approval end points for chronic cutaneous ulcer studies. *Wound Repair and Regeneration* **2012**, *20*, 793–796.

8. World Health Organization Burns Fact Sheet. https://www.who.int/news-room/fact-sheets/detail/burns, 2024; Accessed: 2024-03-19.

9. Sen, C. K. Human wounds and its burden: an updated compendium of estimates. *Advances in Wound Care* **2019**, *8*, 39–48.

10. Church, D.; Elsayed, S.; Reid, O.; Winston, B.; Lindsay, R. Burn wound infections. *Clinical Microbiology Reviews* **2006**, *19*, 403–434.

11. Krishnan, P.; Frew, Q.; Green, A.; Martin, R.; Dziewulski, P. Cause of death and correlation with autopsy findings in burns patients. *Burns* **2013**, *39*, 583–588.

12. Raveendran, P.; Fu, J.; Wallen, S. L. Completely "green" synthesis and stabilization of metal nanoparticles. *Journal of the American Chemical Society* **2003**, *125*, 13940–13941.

13. Bandeira, M.; Chee, B. S.; Frassini, R.; Nugent, M.; Giovanela, M.; Roesch-Ely, M.; Crespo, J. d. S.; Devine, D. M. Antimicrobial PAA/PAH electrospun fiber containing green synthesized zinc oxide nanoparticles for wound healing. *Materials* **2021**, *14*, 2889.

14. White, R. J. An historical overview of the use of silver in wound management. *British Journal of Community Nursing* **2001**, *6*, 3–8.

15. Ahmadi, M.; Adibhesami, M. The effect of silver nanoparticles on wounds contaminated with *Pseudomonas aeruginosa* in mice: an experimental study. *Iranian Journal of Pharmaceutical Research: IJPR* **2017**, *16*, 661–669.

16. Seo, S. B.; Dananjaya, S. H. S.; Nikapitiya, C.; Park, B. K.; Gooneratne, R.; Kim, T. Y.; De Zoysa, M. Silver nanoparticles enhance wound healing in zebrafish (*Danio rerio*) by promoting wound closure. *Fish Shellfish Immunology* **2017**, *68*, 536–545.

17. Vijayakumar, V.; Samal, S. K.; Mohanty, S.; Nayak, S. K. Recent advancements in biopolymer and metal nanoparticle-based materials in diabetic wound healing management. *Inter- national Journal of Biological Macromolecules* **2019**, *122*, 137–148.

18. Kairyte, K.; Kadys, A.; Luksiene, Z. Antibacterial and antifungal activity of photoactivated ZnO nanoparticles in suspension. *Journal of Photochemistry and Photobiology B: Biology* **2013**, *128*, 78–84.

19. Kajbafvala, A.; Ghorbani, H.; Paravar, A.; Samberg, J. P.; Kajbafvala, E.; Sadrnezhaad, S. Effects of morphology on photocatalytic performance of Zinc oxide nanostructures synthe- sized by rapid microwave irradiation methods. *Superlattices and Microstructures* **2012**, *51*, 512–522.

20. Lansdown, A. B.; Mirastschijski, U.; Stubbs, N.; Scanlon, E.; ˚Agren, M. S. Zinc in wound healing: theoretical, experimental, and clinical aspects. *Wound repair and regeneration* **2007**, *15*, 2–16.

21. Chen, S.; Wang, Y.; Bao, S.; Yao, L.; Fu, X.; Yu, Y.; Lyu, H.; Pang, H.; Guo, S.; Zhang, H.; others Cerium oxide nanoparticles in wound care: a review of mechanisms and therapeutic applications. *Frontiers in Bioengineering and Biotechnology* **2024**, *12*, 1404651.

22. Yi, L.; Yu, L.; Chen, S.; Huang, D.; Yang, C.; Deng, H.; Hu, Y.; Wang, H.; Wen, Z.; Wang, Y.; others The regulatory mechanisms of cerium oxide nanoparticles in oxidative stress and emerging applications in refractory wound care. *Frontiers in Pharmacology* **2024**, *15*, 1439960.

23. Allu, I.; Kumar Sahi, A.; Kumari, P.; Sakhile, K.; Sionkowska, A.; Gundu, S. A brief review on cerium oxide (CeO2NPs)-based scaffolds: recent advances in wound healing applications. *Micromachines* **2023**, *14*, 865.

24. Singh, A.; Singh, A. K.; Narayan, G.; Singh, T. B.; Shukla, V. K. Effect of Neem oil and Haridra on non-healing wounds. *AYU (an international quarterly journal of research in Ayurveda)* **2014**, *35*, 398–403.

25. Chundran, N. K.; Husen, I. R.; Rubianti, I. Effect of neem leaves extract (Azadirachta indica) on wound healing. *Althea Medical Journal* **2015**, *2*, 199–203.

26. Jayalakshmi, M. S.; Thenmozhi, P.; Vijayaraghavan, R. Plant leaves extract irrigation on wound healing in diabetic foot ulcers. *Evidence-Based Complementary and Alternative Medicine* **2021**, *2021*, 9924725.

27. Asif, M.; Chaudhry, A. S.; Ashar, A.; Rashid, H. B.; Saleem, M. H.; Aslam, H. B.; Aziz, A. Zinc oxide nanoparticles accelerate the healing of methicillin-resistant *Staphylococcus aureus* (MRSA)-infected wounds in rabbits. *Asian Pacific Journal of Tropical Biomedicine* **2023**, *13*, 488–496.

28. Nosrati, H.; Heydari, M.; Khodaei, M. Cerium oxide nanoparticles: Synthesis methods and applications in wound healing. *Materials Today Bio* **2023**, *23*, 100823.

29. Boonkaew, B. e. a. Agar-based wound dressings with antimicrobial properties. *Carbohydrate Polymers* **2014**, *103*, 456–463.

30. Lee, C. e. a. Fish collagen: Extraction, characterization, and applications for biomaterials. *Marine Drugs* **2020**, *18*, 389.

31. Mali, A. e. a. Neem (Azadirachta indica) extract loaded hydrogel promotes wound healing in diabetic rats. *Journal of Ethnopharmacology* **2020**, *262*, 113219.

32. Chopra, H.; Mohanta, Y. K.; Mahanta, S.; Mohanta, T. K.; Singh, I.; Avula, S. K.; Mallick, S. P.; Rabaan, A. A.; AlSaihati, H.; Alsayyah, A.; others Recent updates in nanotechnological ad-vances for wound healing: A narrative review. *Nanotechnology Reviews* **2023**, *12*, 20230129.

33. Gaikwad, S.; Birla, S.; Ingle, A. P.; Gade, A.; Ingle, P.; Golińska, P.; Rai, M. Superior in vivo wound-healing activity of mycosynthesized silver nanogel on different wound models in rat. *Frontiers in Microbiology* **2022**, *13*, 881404.

34. Al-Nadaf, A. H.; Awadallah, A.; Hussein, R. Superior rat wound-healing activity of green synthesized silver nanoparticles from acetonitrile extract of Juglans regia L.: Pellicle and leaves. *Heliyon* **2024**, *10*, e123456.

35. Dehghani, P.; Akbari, A.; Saadatkish, M.; Varshosaz, J.; Kouhi, M.; Bodaghi, M. Accelera- tion of wound healing in rats by modified lignocellulose-based sponge containing pentoxifylline-loaded lecithin/chitosan nanoparticles. *Gels* **2022**, *8*, 658.